\documentstyle[epsfig]{mn2e} 

\def\etal{{\it et al. }} 
\title[Globular Clusters in NGC~4649] 
{Gemini/GMOS Spectra of Globular Clusters in the Virgo Giant Elliptical NGC~4649} 

\author[Pierce \etal] {
  Michael Pierce$^{1}$\thanks{mpierce@astro.swin.edu.au},
  Terry Bridges$^{2}$, Duncan A. Forbes$^{1}$, Robert Proctor$^{1}$,\\
\\
\LARGE  Michael A. Beasley$^{3}$, Karl Gebhardt$^{4}$, Favio Raul Faifer$^{5,6}$,  Juan Carlos Forte$^{5}$,\\
\\
\LARGE
  Stephen E. Zepf $^{7}$,
  Ray Sharples$^{8}$,
  David A. Hanes$^{2}$\\
\\
  $^1$ Centre for Astrophysics \& Supercomputing, Swinburne University, 
  Hawthorn, VIC 3122, Australia\\ 
  $^2$ Department of Physics, Queen's University, 
  Kingston, ON K7L 3N6, Canada\\
  $^3$ Lick Observatory, University of California, 
  Santa Cruz, CA 95064, USA\\ 
  $^4$ Astronomy Department, University of Texas, 
  Austin, TX 78712, USA\\
  $^5$ Facultad de Cs. Astronomicas y Geofisicas, 
  UNLP, Paseo del Bosque 1900, La Plata, and CONICET, Argentina\\
  $^6$ IALP - CONICET, Argentina\\
  $^7$ Department of Physics and Astronomy, Michigan State University, 
  East Lansing, MI 48824, USA\\
  $^8$ Department of Physics, University of Durham, 
  South Road, Durham DH1 3LE\\
}

\begin{document} 
\maketitle 

\begin{abstract} 

NGC~4649 (M60) is one of a handful of giant Virgo ellipticals. We have
obtained Gemini/GMOS spectra for 38 GCs associated with this galaxy.
Applying the multi-index $\chi^2$ minimisation technique of Proctor \&
Sansom (2002) with the single stellar population models of Thomas,
Maraston \& Korn (2004) we derive ages, metallicities and
$\alpha$-element abundance ratios.  We find several young (2--3 Gyr
old) super-solar metallicity GCs, while the majority are old ($>$10
Gyrs), spanning a range of metallicities from solar to [Z/H]=--2. At
least two of these young GCs are at large projected radii of 17-20
kpc. The galaxy itself shows no obvious signs of a recent starburst,
interaction or merger. A trend of decreasing $\alpha$-element ratio
with increasing metallicity is found.

\end{abstract} 
 

\section{Introduction}\label{sec_intro} 

Accurately determined ages, metallicities and $\alpha$-element
abundances of globular cluster (GC) systems across the entire range of
galaxy types can provide strong constraints for galaxy formation
models.  One example of such a constraint is the recently observed
trend of decreasing $\alpha$/Fe ratios with increasing metallicity for
extra-galactic GC systems (Puzia \etal 2005; Pierce \etal
2005a,b). This trend restricts the allowable chemical enrichment
histories, relative contribution of Type Ia vs Type II supernova, and
star formation timescales for galaxies well beyond the Local Group,
where there is little possibility of directly resolving stellar
populations.

To accurately measure ages, metallicities and abundance ratios with
Lick indices from low-resolution spectra requires a minimum S/N of
$\sim$30, which corresponds to an H$\beta$ error of $\pm$0.3\AA\
(Cardiel \etal 2003).  With an integration time of $\sim$8 hours on an
8-metre class telescope it is possible to obtain multi-object spectra
to this depth for the brightest GCs of a rich GC system for galaxies
within $\sim$20 Mpc of the Milky Way.

A major aim of most moderate to high S/N GC spectroscopy has been to
measure ages and hence infer when GC formation occurred. Spectroscopic
follow-up is a complimentary approach to recent photometric
results. Rhode, Zepf \& Santos (2005) show that the mass-normalised
number of blue GCs increases with host galaxy mass.  This suggests
that the formation of blue GCs is affected by the mass of the host
halo.  The observed ``blue tilt'' correlation between GC luminosity
and colour of the blue, metal-poor, sub-population (Strader \etal
2005; Harris \etal 2005) requires significant spectroscopic follow-up
to be both confirmed and understood. Another property of the blue
sub-population to be studied is the correlation between host galaxy
luminosity and mean blue GC sub-population colour (Strader, Brodie \&
Forbes 2004b), which implies a correlation between galaxy mass and
blue GC metallicity.  This is another indication that the formation of
blue GCs is affected by the mass of the host halo.

The majority of the most recent GC spectroscopy for large ellipticals
has focused on group ellipticals (e.g. NGC 1052 Pierce \etal 2005a;
NGC 3379 Pierce \etal 2005b; NGC 3610 Strader \etal 2003,2004a; NGC
5128 Peng \etal 2004; NGC 2434, NGC 3379, NGC 3585, NGC 5846 and NGC
7192 Puzia \etal 2004).

From the literature there are several large ellipticals in clusters
for which GC spectra have been analysed to measure ages and
metallicities. Cohen, Blakeslee \& Ryzhov (1998) present spectral
indices for 150 of M87's GCs.  These vary in S/N, with a sizeable
fraction of high enough quality to be useful. Co-adding GCs of similar
metallicity, they find the GCs are generally old ($\geq$10 Gyrs) with
metallicities spanning from [Fe/H]=--2 to above solar.  Cohen,
Blakeslee \& Cote (2003) measure metallicities for 47 GCs associated
with NGC~4472~(M49).  However the S/N is less than 30 for all of the
spectra and therefore ages for individual GCs cannot be measured with
any confidence.  Beasley \etal (2000) co-added spectra for NGC~4472
GCs finding both the metal-rich and metal-poor sub-populations to be
old.

For the Fornax cluster, Kissler-Patig \etal (1998) measured
metallicities for 18 GCs around NGC~1399. Forbes \etal (2001)
presented higher S/N spectra for 10 GCs and found two to be young
(1-2~Gyrs). From the Goudfrooij \etal (2001) spectroscopic sample of NGC~1316,
only the spectra of 3 exceptionally bright GCs have high enough S/N to
measure ages.  The Lick indices of all 3 bright GCs indicate a young
age ($\sim$3~Gyrs) corresponding to the host galaxy's recent merger
event.  The GC colour-magnitude plot for NGC~1316 is atypical in that
there is a significant population of bright, intermediate-colour GCs.

One galaxy with what appears to be conflicting results is NGC~4365 in
the Virgo cluster. Brodie \etal (2005) present spectroscopic results
for NGC~4365 GCs and find some GCs previously thought to belong to an
intermediate age sub-population (Larsen \etal 2003; Puzia \etal 2002),
are in fact an intermediate metallicity sub-population with old ages.
Kundu \etal (2005) present HST NIC3 H-band data and find that these
agree with earlier claims of a GC sub-population with intermediate
ages 2-8 Gyrs. There appears to be no consensus yet for the GC
population of this galaxy, unlike other systems.

While possessing a larger GC population, cD and brightest cluster
galaxies have the additional complication of GCs potentially
associated with the cluster potential.  The Virgo cluster elliptical
NGC~4649 (M60) is a worthy target for GC spectroscopy as a
non-central, giant, cluster elliptical.

NGC 4649 (M60) is luminous, M$_V$=--22.38, and relatively nearby at
D=16.8 Mpc.  UV data for the galaxy light suggests a major old
population, plus minor on-going star-formation (Magris \& Bruzual
1993). However, from optical spectra, Terlevich \& Forbes (2002)
measure Lick indices and obtain an age of 11 Gyr, metallicity of
[Fe/H]=+0.3 and [Mg/Fe]=+0.3 for the central regions of the
galaxy. The Chandra X-ray observations of Randall \etal (2004) find
some structure in the diffuse gas.

The GC imaging study of Forbes \etal (2004) found the standard colour
bimodality and S$_N$=4.1.  Assuming the two sub-populations have
similar mass functions, this suggests a similar formation
age. NGC~4649 is one of the galaxies with the observed ``blue-tilt''
(Strader \etal 2005). Recently Spitler \etal (2006) have shown that
the ``blue-tilt'' is present amongst the GC population of a nearby
spiral galaxy, the Sombrero.  Unfortunately our spectroscopic sample
does not sample far enough down the luminosity function to adequately
test hypotheses regarding the ``blue-tilt''.

\section{Observations and data reduction}\label{sec_obs} 

The observations described below are part of Gemini program
GN-2002A-Q-13.  GC candidates were selected from Gemini North
Multi-Object Spectrograph (GMOS; Hook \etal 2002) imaging, obtained
during 2002 April, for three fields around NGC~4649. The data
reduction and GC candidate selection process is described in Forbes
\etal (2004) and Bridges \etal (2006).  

GMOS masks for three fields were designed, but only the central field
was observed within the time allocated. Spectra of NGC~4649
globular clusters were obtained with GMOS on the Gemini North
telescope during 2003 on May 31, June 1, June 4 and June 27.  Seeing
ranged from 0.65--0.9 arc-seconds over the four nights. Exposures of
16$\times$1800s were taken, yielding a total of 8 hours on-source
integration time.  Bias frames, dome flat-fields and Copper-Argon
(CuAr) arc exposures were taken as part of the Gemini baseline
calibrations.  From the CuAr arcs, wavelength solutions with typical
residuals of 0.1\AA\ were achieved.

These data were reduced using the Gemini/GMOS packages in IRAF and a
number of custom made scripts (see Bridges \etal 2006 for
details). After some experimentation, optimal (variance) extraction
was found to yield the best results since our data are over-sampled on
the detector. In some cases, objects were too faint to trace
individually and we therefore co-added several 2-d images, taken
adjacent in time, to act as a reference for the extractions. We
verified beforehand that flexure was minimal between the reference
images.  Finally, the extracted spectra were median combined and
weighted by their fluxes with cosmic ray rejection.

In the absence of any velocity standard stars, the recession
velocities were measured by using six Bruzual \& Charlot (2003) model
stellar energy distributions (SEDs) for 14 and 5 Gyr ages with
metallicities [Fe/H] = --1.64, --0.33 and +0.1. The task FXCOR in IRAF
was used and the average was taken.  Objects with recession velocities
in the range 1100$\pm$600 km/s are potentially associated with
NGC~4649.  These are presented in Table \ref{tabobs}. There was one
background object (a QSO at z$\sim$0.5) out of the 39 spectra
obtained.  Our low contamination rate of 2.5\% is due to good imaging
and colour selection.

In order to measure Lick indices, we convolved our spectra with a
wavelength-dependent Gaussian kernel to match the resolution of the
Lick/IDS system (see Beasley \etal 2004b). Lick indices (Trager \etal
1998) were measured from our spectra.  Due to the variable wavelength
ranges in these spectra, the same set of indices could not be measured
for all spectra. However, all covered a wavelength range of
4500--5500~\AA. Uncertainties in the indices were derived from the
photon noise in the unfluxed spectra.  No Lick standard stars were
observed so we therefore cannot fully calibrate the GCs onto the Lick
system (see Pierce \etal 2005b).  Consequently there are some small
systematic differences between some of the measured indices and those
used in the SSP models.  These issues are discussed further in
Sections \ref{sec_age} and \ref{sec_disc}. Measured line indices and
are presented in Tables \ref{tabindex1} and \ref{tabindex2}.  Line
index uncertainties are presented in Tables \ref{tabinderr1} and
\ref{tabinderr2}. The final spectra have S/N = 5--21
\AA$^{-1}$ at 5000 \AA, giving errors in the H$\beta$ index of
0.12--0.38 \AA.

\begin{table*} 
 \begin{center}
  \caption{\scriptsize Confirmed globular clusters around NGC~4649.  Cluster ID, coordinates, {\it g} magnitude and {\it g--i} colours are from our GMOS imaging and are instrumental magnitudes only.  Heliocentric velocities are from the spectra presented in this work.}
\renewcommand{\arraystretch}{1.5} 
\begin{tabular}{lcccccccc} 
\hline 
\hline 
ID & R.A. & Dec. & X & Y & g & g--i & V$_{helio}$ & V$_{err}$ \\
   & (J2000) & (J2000) & (pixels) & (pixels) & (mag) & (mag) & (km/sec) & (km/sec) \\
\hline
  89 & 190.983978 &  11.536740 & 1324.96 &  478.67 &  22.72 &  1.08 & 1199.2 &   43.4 \\
 124 & 190.981033 &  11.536771 & 1339.40 &  621.02 &  21.91 &  0.88 & 833.4 &   59.6 \\
  68 & 190.986160 &  11.539418 & 1447.41 &  362.65 &  22.58 &  1.22 & 649.3 &   33.6 \\
 148 & 190.977036 &  11.545153 & 1769.64 &  777.98 &  22.74 &  1.14 & 1184.8 &   46.8 \\
 175 & 190.974426 &  11.531744 & 1120.73 &  961.31 &  22.29 &  1.18 & 852.3 &   37.6 \\
 183 & 190.972626 &  11.538513 & 1461.97 & 1019.24 &  22.26 &  1.13 & 1299.4 &   35.8 \\
 158 & 190.971558 &  11.589099 & 3957.80 &  853.95 &  23.21 &  1.16 & 1261.0 &   46.0 \\
 360 & 190.949738 &  11.594447 & 4316.58 & 1883.79 &  21.20 &  0.91 & 1270.1 &   40.1 \\
 329 & 190.957642 &  11.538593 & 1531.35 & 1741.23 &  21.93 &  0.88 & 1511.2 &   65.3 \\
 277 & 190.962753 &  11.548118 & 1978.06 & 1454.03 &  22.59 &  1.04 & 1299.1 &   46.5 \\
 251 & 190.960129 &  11.599953 & 4542.24 & 1358.58 &  22.29 &  1.09 & 1099.7 &   43.9 \\
 298 & 190.960342 &  11.551459 & 2153.17 & 1556.03 &  22.58 &  0.90 & 1063.0 &   52.5 \\
 318 & 190.959274 &  11.536565 & 1424.30 & 1671.14 &  23.08 &  1.00 & 1173.4 &   48.6 \\
 606 & 190.936737 &  11.548500 & 2110.58 & 2706.89 &  21.36 &  1.18 & 627.0 &   39.6 \\
 558 & 190.938461 &  11.555140 & 2430.02 & 2595.52 &  21.75 &  0.74 & 1087.8 &   50.8 \\
 434 & 190.949310 &  11.528630 & 1077.12 & 2185.82 &  22.18 &  1.04 & 925.9 &   36.6 \\
 462 & 190.947189 &  11.532343 & 1269.20 & 2272.22 &  22.18 &  0.92 & 1112.8 &   45.4 \\
 517 & 190.942352 &  11.539107 & 1623.47 & 2476.30 &  22.03 &  1.14 & 857.2 &   56.3 \\
 412 & 190.945847 &  11.582625 & 3751.32 & 2121.39 &  22.48 &  0.95 & 483.6 &   47.5 \\
 502 & 190.942627 &  11.559628 & 2632.89 & 2375.35 &  22.10 &  1.17 & 688.6 &   43.5 \\
 640 & 190.931168 &  11.586047 & 3984.06 & 2815.36 &  22.61 &  0.79 & 1015.5 &  104.5 \\
 740 & 190.932129 &  11.529132 & 1176.86 & 3012.11 &  21.68 &  0.98 & 962.1 &   44.8 \\
 806 & 190.929291 &  11.538971 & 1673.81 & 3107.00 &  21.50 &  1.18 & 1197.4 &   46.2 \\
 899 & 190.927261 &  11.533805 & 1428.31 & 3227.10 &  21.44 &  0.92 & 1372.0 &   42.7 \\
 975 & 190.924591 &  11.539021 & 1696.82 & 3333.25 &  20.99 &  0.92 & 1052.2 &   42.3 \\
1063 & 190.920807 &  11.535969 & 1563.10 & 3529.46 &  21.37 &  0.94 & 953.9 &   47.3 \\
1011 & 190.918243 &  11.596663 & 4563.36 & 3393.26 &  21.95 &  0.94 & 1135.0 &   45.7 \\
1037 & 190.919540 &  11.564267 & 2962.26 & 3469.39 &  21.97 &  1.13 &  1017.1 &   42.6 \\
1145 & 190.916580 &  11.532989 & 1434.77 & 3745.73 &  21.60 &  1.08 &  926.8 &   47.3 \\ 
1252 & 190.910614 &  11.544307 & 2018.25 & 3985.40 &  21.27 &  0.89 &  1134.2 &   47.0 \\
1384 & 190.906281 &  11.550952 & 2364.45 & 4165.70 &  20.91 &  1.03 &  1122.0 &   58.2 \\
1126 & 190.918457 &  11.526026 & 1083.67 & 3685.07 &  22.21 &  0.90 &  1346.1 &   47.9 \\
1298 & 190.909515 &  11.536103 & 1619.00 & 4073.00 &  22.25 &  1.13 &  1324.5 &   38.0 \\
1211 & 190.912033 &  11.544750 & 2033.87 & 3914.83 &  22.29 &  1.01 &  889.3 &   69.3 \\ 
1098 & 190.914841 &  11.581110 & 3812.20 & 3623.63 &  22.27 &  0.77 &  1305.2 &   39.3 \\
1182 & 190.911743 &  11.562973 & 2932.52 & 3850.43 &  22.32 &  1.15 &  1454.0 &   45.1 \\
1574 & 190.896591 &  11.602605 & 4950.56 & 4411.99 &  21.65 &  1.01 &  703.0 &   53.9 \\ 
1443 & 190.904984 &  11.551611 & 2402.56 & 4225.51 &  22.65 &  1.28 &  667.9 &   63.4 \\ 
\hline
\hline 
\label{tabobs}
\end{tabular} 
\end{center} 
\end{table*}

\begin{table*} 
\begin{center}
\caption{\scriptsize Globular cluster indices $\lambda <$ 4600 \AA.  Indices in brackets are removed during the fitting process.   Missing values are due to limited wavelength coverage. Index errors are presented in Table 3.}
\renewcommand{\arraystretch}{1.5}
\begin{tabular}{lcccccccccccc}
\hline 
\hline 
ID&H$\delta_{A}$&H$\delta_{F}$&CN$_{1}$&CN$_{2}$&Ca4227&G band&H$\gamma_{A}$&H$\gamma_{F}$&Fe4383&Ca4455& Fe4531 \\
     & (\AA) & (\AA) & (mag) & (mag) & (\AA) & (\AA) & (\AA) & (\AA) &  (\AA) & (\AA) & (\AA) \\
\hline
  89 & (-8.936)&(-0.154)& (0.125)& (0.165)&  0.448 &  5.290 & -1.691 &  0.944 &  2.259 &  1.085 &  1.939 \\
 124 &   3.226 &  2.950 &(-0.008)& (0.035)&  0.621 & (3.662)&  1.155 &  2.004 &  1.883 &  0.008 &  1.310 \\ 
  68 &  (2.165)& (3.963)& (0.033)& (0.101)&  1.045 &  5.253 & -5.858 & -0.766 & (8.650)&  2.139 & (5.030)\\ 
 148 &(-10.481)&(-3.591)& (0.295)& (0.357)&  1.440 & (2.541)&(-3.292)& (0.473)&  6.518 &  0.823 &  3.553 \\
 175 & (-2.707)&(-0.900)& (0.168)& (0.173)&  1.016 &  2.234 & -2.255 & -0.180 &  2.729 &  2.310 &  3.974 \\ 
 183 &   0.270 &  0.831 & (0.083)& (0.051)& -0.004 &  5.785 & -3.500 & -0.165 &  4.196 &  1.225 &  2.247 \\
 158 &  -3.139 & -0.488 & (0.180)& (0.151)&  0.882 &  6.213 & -1.419 &  0.893 &  3.112 &  1.406 &  3.244 \\
 360 &  ...... & ...... & ...... & ...... &  0.865 &  3.528 & -1.488 &  0.556 &  2.904 &  1.068 &  2.187 \\
 329 &   1.989 &  1.520 &(-0.085)&(-0.078)&  0.305 & (1.784)& -0.260 &  0.830 &  2.898 &  0.702 &  0.928 \\
 277 &  (0.415)& (2.965)&(-0.070)&(-0.041)&  0.614 &  4.777 & -6.229 & -2.614 &  4.175 &  0.609 &  2.223 \\
 251 &  ...... & ...... & ...... & ...... & ...... &  6.561 & -4.919 & -1.155 &  2.320 & (0.087)&  2.617 \\
 298 &   3.471 &(-1.123)&(-0.137)&(-0.142)&(-1.050)&  0.375 & (3.654)& (3.162)&  1.359 & (1.426)&  0.426 \\
 318 & (-6.529)&(-2.985)& (0.181)& (0.190)&  1.187 &  4.917 &  0.321 &  1.237 & -1.053 &  0.943 & (6.004)\\
 606 &  -0.875 &  0.697 & (0.145)& (0.197)& (0.401)&  5.062 & -5.650 & -1.572 &  4.103 &  1.209 &  3.229 \\ 
 558 &   3.361 &  2.874 &(-0.061)&(-0.030)&  0.183 & (3.782)&  1.268 & (3.135)& (3.474)&  0.382 &  1.311 \\
 434 &  -0.600 & -0.655 & (0.027)& (0.060)&  0.182 &  6.155 & -4.006 & -1.522 & (0.478)&  0.618 &  2.909 \\ 
 462 &   0.503 &  2.451 & (0.096)& (0.122)&  1.444 &  5.113 & -3.244 & -0.350 &  0.392 &  0.793 &  2.908 \\
 517 &  (4.415)& (4.534)& (0.043)& (0.033)&(-0.051)&  4.880 & -3.174 & -0.156 &  2.425 &  1.193 &  2.759 \\ 
 412 &   0.235 &  0.859 &(-0.054)&(-0.072)&(-0.867)&  4.880 & -3.487 &  0.392 &  2.514 &  1.509 &  1.864 \\ 
 502 & (-4.873)&(-1.419)& (0.209)& (0.225)&  1.121 &  5.483 & -4.837 & -0.666 &  4.282 & (0.933)&  4.245 \\ 
 640 &  (1.353)& (1.588)& (0.035)& (0.052)& (2.221)&  0.728 & (6.237)&  3.149 & -4.268 &  1.415 &  0.314 \\
 740 &   2.822 &  3.308 & (0.011)& (0.027)&  0.617 &  3.172 & -0.098 &  1.478 &  2.720 &  0.625 &  1.785 \\ 
 806 & (-5.422)&(-1.911)& (0.302)& (0.361)&  1.077 & (3.982)& -6.102 & -0.699 &  6.163 &  1.663 &  3.206 \\
 899 &   1.250 &  1.462 & (0.024)& (0.068)&  0.831 &  4.754 & -2.170 &  0.321 &  2.233 &  0.133 &  2.367 \\
 975 &   3.233 &  2.390 &(-0.030)& (0.022)&  0.478 &  3.632 & -0.767 &  0.980 &  1.782 &  0.150 &  2.049 \\
1063 &   0.349 &  1.798 & (0.028)& (0.057)&  0.076 &  2.937 & -1.370 &  1.176 &  3.206 &  0.733 &  1.619 \\ 
1011 &  ...... & ...... & ...... & ...... & ...... & (5.206)& -1.727 & (1.520)&  3.486 &(-0.183)&  2.550 \\
1037 &  -0.260 & -1.406 & (0.148)& (0.202)& (0.405)& (1.864)&(-2.044)& -0.808 &  3.080 &  0.460 &  4.094 \\
1145 &  -0.414 &  0.605 & (0.083)& (0.107)&  0.635 &  5.077 & -3.217 & -0.668 &  2.842 &  1.186 &  2.451 \\ 
1252 &   1.541 &  2.265 & (0.048)& (0.076)&  0.694 &  3.673 & -1.594 &  0.725 &  2.029 &  0.125 &  1.823 \\
1384 & (-0.236)& (1.215)& (0.068)& (0.092)&  0.618 &  3.596 & -1.039 &  0.727 &  1.054 &  0.569 &  2.381 \\
1126 & (-1.003)& (1.215)& (0.078)& (0.105)&  0.032 & (2.886)&  1.166 &  2.330 & (5.764)&  1.223 &  0.288 \\
1298 & (-3.213)& (1.650)& (0.131)& (0.179)&  0.643 &  5.052 & -3.889 & -0.447 &  4.607 &  1.422 &  2.993 \\
1211 & (-2.832)& (0.173)& (0.222)& (0.245)&  1.118 &  4.989 & -2.369 &  0.862 &  2.320 &  0.751 &  1.741 \\ 
1098 &   2.101 &  1.964 & (0.002)& (0.018)&(-1.346)&  3.909 &  1.203 &  1.692 & -0.399 & -0.643 &  1.470 \\
1182 &  -1.099 &  0.500 & (0.204)& (0.231)&  0.705 &  3.754 & -4.786 & -0.178 &  5.309 &  0.547 &  1.187 \\
1574 &  ...... & ...... & ...... & ...... & ...... & ...... & ...... & ...... & ...... &  0.443 &  2.740 \\ 
1443 & (-8.897)&(-2.203)& (0.250)& (0.350)&  1.441 &  3.752 & -4.720 & -0.683 &  5.693 &(-1.506)& (0.000)\\ 
\hline 
\label{tabindex1}
\end{tabular} 
\end{center} 
\end{table*} 

\begin{table*} 
\begin{center}
\caption{\scriptsize Globular cluster indices errors $\lambda <$4600 \AA.  Missing values are due to limited wavelength coverage. Index errors are derived from photon noise in the unfluxed spectra.}
\renewcommand{\arraystretch}{1.5}
\begin{tabular}{lcccccccccccc}
\hline 
\hline 
ID&H$\delta_{A}$&H$\delta_{F}$&CN$_{1}$&CN$_{2}$&Ca4227&G band&H$\gamma_{A}$&H$\gamma_{F}$&Fe4383&Ca4455& Fe4531 \\
     & (\AA) & (\AA) & (mag) & (mag) & (\AA) & (\AA) & (\AA) & (\AA) &  (\AA) & (\AA) & (\AA) \\
\hline
  89 &   0.822 &  0.472 &  0.017 &  0.020 &  0.322 &  0.500 &  0.539 &  0.328 &  0.731 &  0.356 &  0.530\\
 124 &   0.375 &  0.253 &  0.011 &  0.013 &  0.204 &  0.341 &  0.336 &  0.208 &  0.494 &  0.252 &  0.368\\ 
  68 &   0.594 &  0.365 &  0.017 &  0.019 &  0.305 &  0.492 &  0.572 &  0.345 &  0.616 &  0.326 &  0.466\\ 
 148 &   0.851 &  0.554 &  0.018 &  0.021 &  0.309 &  0.551 &  0.566 &  0.336 &  0.686 &  0.369 &  0.519\\
 175 &   0.606 &  0.413 &  0.015 &  0.018 &  0.261 &  0.472 &  0.458 &  0.293 &  0.610 &  0.281 &  0.420\\ 
 183 &   0.515 &  0.353 &  0.014 &  0.016 &  0.271 &  0.399 &  0.455 &  0.281 &  0.571 &  0.284 &  0.418\\
 158 &   0.989 &  0.649 &  0.024 &  0.028 &  0.436 &  0.656 &  0.729 &  0.459 &  0.971 &  0.478 &  0.667\\
 360 &  ...... & ...... & ...... & ...... &  0.160 &  0.273 &  0.281 &  0.175 &  0.383 &  0.188 &  0.280\\
 329 &   0.397 &  0.279 &  0.011 &  0.013 &  0.212 &  0.367 &  0.351 &  0.223 &  0.494 &  0.245 &  0.371\\
 277 &   0.597 &  0.368 &  0.016 &  0.019 &  0.304 &  0.505 &  0.579 &  0.381 &  0.692 &  0.353 &  0.515\\
 251 &  ...... & ...... & ...... & ...... & ...... &  0.433 &  0.524 &  0.331 &  0.673 &  0.332 &  0.459\\
 298 &   0.499 &  0.421 &  0.015 &  0.017 &  0.321 &  0.493 &  0.429 &  0.263 &  0.692 &  0.325 &  0.514\\
 318 &   0.730 &  0.504 &  0.018 &  0.021 &  0.325 &  0.562 &  0.582 &  0.366 &  0.888 &  0.428 &  0.570\\
 606 &   0.384 &  0.257 &  0.010 &  0.012 &  0.181 &  0.284 &  0.323 &  0.204 &  0.389 &  0.193 &  0.277\\ 
 558 &   0.364 &  0.246 &  0.011 &  0.012 &  0.201 &  0.328 &  0.321 &  0.187 &  0.454 &  0.236 &  0.351\\
 434 &   0.514 &  0.378 &  0.013 &  0.015 &  0.248 &  0.371 &  0.438 &  0.286 &  0.599 &  0.286 &  0.403\\ 
 462 &   0.488 &  0.310 &  0.013 &  0.015 &  0.220 &  0.383 &  0.424 &  0.267 &  0.581 &  0.279 &  0.406\\
 517 &   0.496 &  0.312 &  0.014 &  0.016 &  0.267 &  0.387 &  0.428 &  0.266 &  0.569 &  0.278 &  0.394\\ 
 412 &   0.561 &  0.386 &  0.015 &  0.018 &  0.318 &  0.460 &  0.519 &  0.312 &  0.690 &  0.328 &  0.485\\ 
 502 &   0.625 &  0.414 &  0.014 &  0.017 &  0.248 &  0.401 &  0.458 &  0.283 &  0.557 &  0.278 &  0.391\\ 
 640 &   0.579 &  0.400 &  0.016 &  0.018 &  0.263 &  0.522 &  0.424 &  0.279 &  0.798 &  0.351 &  0.560\\
 740 &   0.362 &  0.235 &  0.010 &  0.012 &  0.194 &  0.318 &  0.317 &  0.195 &  0.448 &  0.228 &  0.332\\ 
 806 &   0.469 &  0.313 &  0.011 &  0.013 &  0.187 &  0.309 &  0.346 &  0.206 &  0.399 &  0.206 &  0.302\\
 899 &   0.336 &  0.231 &  0.009 &  0.011 &  0.167 &  0.276 &  0.298 &  0.186 &  0.404 &  0.204 &  0.291\\
 975 &   0.257 &  0.179 &  0.007 &  0.009 &  0.137 &  0.226 &  0.232 &  0.144 &  0.331 &  0.169 &  0.242\\
1063 &   0.333 &  0.219 &  0.009 &  0.010 &  0.170 &  0.278 &  0.280 &  0.169 &  0.382 &  0.194 &  0.287\\ 
1011 &  ...... & ...... & ...... & ...... & ...... &  0.366 &  0.404 &  0.238 &  0.540 &  0.284 &  0.392\\
1037 &   0.503 &  0.381 &  0.013 &  0.015 &  0.245 &  0.412 &  0.405 &  0.264 &  0.542 &  0.272 &  0.370\\
1145 &   0.401 &  0.274 &  0.010 &  0.012 &  0.189 &  0.302 &  0.334 &  0.214 &  0.435 &  0.212 &  0.313\\ 
1252 &   0.302 &  0.200 &  0.008 &  0.010 &  0.151 &  0.255 &  0.266 &  0.165 &  0.372 &  0.191 &  0.276\\
1384 &   0.276 &  0.184 &  0.007 &  0.009 &  0.135 &  0.227 &  0.228 &  0.143 &  0.324 &  0.159 &  0.231\\
1126 &   0.491 &  0.321 &  0.013 &  0.015 &  0.248 &  0.403 &  0.387 &  0.234 &  0.521 &  0.273 &  0.429\\
1298 &   0.572 &  0.340 &  0.014 &  0.016 &  0.259 &  0.410 &  0.458 &  0.285 &  0.568 &  0.283 &  0.421\\
1211 &   0.574 &  0.376 &  0.014 &  0.016 &  0.243 &  0.403 &  0.441 &  0.267 &  0.595 &  0.295 &  0.438\\ 
1098 &   0.473 &  0.323 &  0.013 &  0.016 &  0.282 &  0.416 &  0.407 &  0.255 &  0.627 &  0.326 &  0.460\\
1182 &   0.562 &  0.376 &  0.015 &  0.017 &  0.262 &  0.441 &  0.479 &  0.289 &  0.579 &  0.310 &  0.437\\
1574 &  ...... & ...... & ...... & ...... & ...... & ...... & ...... & ...... & ...... &  0.230 &  0.327\\ 
1443 &   0.828 &  0.524 &  0.017 &  0.020 &  0.298 &  0.520 &  0.560 &  0.347 &  0.667 &  0.389 &  0.532\\ 
\hline 
\label{tabinderr1}
\end{tabular} 
\end{center} 
\end{table*}

\begin{table*} 
\begin{center} 
\caption{\scriptsize Globular cluster indices $\lambda >$4600 \AA.  Indices in brackets are removed during the fitting process.  Missing values are due to limited wavelength coverage. Index errors are presented in Table 5.}
\renewcommand{\arraystretch}{1.5}
\begin{tabular}{lccccccccccc} 
\hline 
\hline 
ID  &  C4668 & H$\beta$ & Fe5015 & Mg$_{1}$ & Mg$_{2}$ &  Mgb   & Fe5270 & Fe5335 & Fe5406 & Fe5709 & Fe5782 \\
   & (\AA)  & (\AA)    & (\AA)  & (mag)    & (mag)    & (\AA)  & (\AA)  & (\AA)  & (\AA)  & (\AA)  & (\AA) \\
\hline 
  89 &   3.089 & (0.060)&  6.110 & (0.022) & (0.140) &  3.582 &  1.318 &  2.239 &  1.677 & ...... & ......  \\
 124 &  -0.554 &  2.540 &  2.414 &(-0.023) & (0.055) &  1.740 &  1.433 &  1.686 &  0.771 & ...... & ......  \\
  68 &   3.689 &  1.188 & (2.959)& (0.030) & (0.207) &  3.641 &  2.867 & (3.201)&  1.719 & ...... & ......  \\
 148 &   4.343 &  1.559 &  5.026 & (0.021) & (0.125) &  3.140 &  2.632 &  1.258 &  1.386 & ...... & ......  \\
 175 &   6.261 &  1.336 &  4.277 & (0.079) & (0.209) &  3.148 &  2.794 &  2.074 &  1.612 & ...... & ......  \\
 183 &   0.592 &  2.337 &  4.266 & (0.021) & (0.149) &  3.265 &  2.208 &  1.830 &  1.233 & ...... & ......  \\
 158 &   2.391 &  0.326 &  3.733 & (0.010) & (0.147) &  3.022 &  2.638 &  1.438 &  1.376 &  0.823 &  1.039  \\
 360 &   2.034 &  2.574 &  3.346 & (0.011) & (0.093) &  2.057 &  1.726 &  1.386 &  1.017 &  0.549 &  0.248  \\
 329 &   1.044 &  2.401 &  3.422 &(-0.018) & (0.059) &  1.844 &  1.085 &  0.730 &  0.342 & ...... & ......  \\
 277 &   1.967 &  0.693 &  3.140 & (0.047) & (0.148) &  3.575 &  2.039 &  1.968 &  1.142 & ...... & ......  \\
 251 &   3.537 &  1.213 &  5.896 & (0.060) & (0.221) &  4.205 &  1.952 &  2.026 &  1.621 &  0.600 &  0.666  \\
 298 &  -0.013 &  2.302 &  2.896 & (0.019) & (0.037) &  1.072 & (1.647)&  1.052 &  1.083 & ...... & ......  \\
 318 & (-5.908)&  0.490 & -1.582 &(-0.032) & (0.093) &  2.665 & -0.015 &  2.098 &  1.380 & ...... & ......  \\
 606 &   4.089 &  1.486 &  4.178 & (0.102) & (0.242) &  3.531 & (1.912)& (2.670)& (1.665)& ...... & ......  \\
 558 &  -1.369 &  2.326 &  2.665 & (0.010) & (0.044) & (0.671)&  1.889 &  1.503 &  0.607 & ...... & ......  \\
 434 &   2.491 &  1.878 &  2.343 & (0.025) & (0.123) &  2.506 &  2.118 &  1.852 &  0.903 & ...... & ......  \\
 462 &   2.792 &  1.203 & (2.264)& (0.023) & (0.104) &  2.234 &  2.104 &  0.806 &  0.988 & ...... & ......  \\
 517 &   4.152 & (1.208)&  4.771 & (0.080) & (0.201) &  3.094 & (3.386)&  1.581 &  1.391 & ...... & ......  \\
 412 &   3.400 & (0.110)&  3.138 & (0.041) & (0.106) &  2.258 & (0.861)&  2.101 &  1.017 &  0.451 &  0.812  \\
 502 &   7.967 &  2.025 &  6.563 & (0.107) & (0.287) &  5.055 &  2.396 &  3.027 &  1.513 &  0.555 & ......  \\
 640 &   2.724 &  2.310 &  0.315 &(-0.055) &(-0.067) &(-0.753)& -0.508 &  1.176 &(-0.343)&  0.189 & (0.752) \\
 740 &   0.912 &  1.340 & (4.897)& (0.010) & (0.109) &  2.759 &  1.770 &  1.713 &  0.935 & ...... & ......  \\
 806 &   4.799 &  0.995 & (2.748)& (0.076) & (0.282) & (5.771)&  2.819 &  2.265 &  1.744 & ...... & ......  \\
 899 &   1.132 &  1.847 &  2.220 & (0.002) & (0.090) &  2.692 &  1.416 & (1.844)&  0.903 & ...... & ......  \\
 975 &   0.648 &  2.222 &  3.512 &(-0.004) & (0.083) &  2.330 &  1.691 &  1.420 &  0.711 & ...... & ......  \\
1063 &   2.715 &  2.109 &  3.420 &(-0.008) & (0.086) &  1.769 &  1.693 &  1.530 &  0.959 & ...... & ......  \\
1011 &   3.098 &  1.602 &  2.698 & (0.072) & (0.135) & (1.685)&  1.984 &  0.972 &  1.159 &  0.508 &  0.430  \\
1037 &  (5.808)&  1.082 &  5.031 & (0.078) & (0.233) &  3.307 &  2.351 &  1.509 &(-0.189)&(-0.008)&  0.410  \\
1145 &   3.328 &  1.800 &  5.140 & (0.041) & (0.182) &  3.746 &  2.080 &  1.744 &  1.449 & ...... & ......  \\
1252 &   0.854 &  2.480 &  4.032 &(-0.016) & (0.084) & (1.908)&  1.484 &  1.480 &  0.726 & ...... & ......  \\
1384 &  (4.074)&  1.862 &  2.771 & (0.059) & (0.150) &  2.590 &  1.839 &  1.390 & (1.181)& ...... & ......  \\
1126 &   0.487 &  2.999 &  1.284 &(-0.013) & (0.065) &  1.288 &  0.919 &  0.772 &  0.099 & ...... & ......  \\
1298 &   1.132 &  1.634 &  2.884 & (0.014) & (0.141) &  3.575 &  1.882 &  1.781 &  0.972 & ...... & ......  \\
1211 &   1.910 &  1.433 &  2.478 & (0.036) & (0.192) & (2.128)&  2.149 & (2.962)& (1.666)& ...... & ......  \\
1098 &  -1.524 & (0.813)&  2.016 &(-0.038) & (0.041) &  2.144 &  1.139 &  0.298 & -0.282 &  0.883 &  0.180  \\
1182 &   5.486 &  1.656 &  5.991 & (0.103) & (0.257) & (5.120)&  1.632 &  2.840 &  1.048 & (1.795)&  0.985  \\
1574 &   0.993 &  1.899 &  3.024 & (0.031) & (0.109) & (2.388)&  1.254 &  0.543 &  1.067 &  0.395 &  0.328  \\
1443 &   6.813 &  1.976 & (8.367)& (0.028) & (0.217) & (5.264)& (0.929)& (1.110)&  2.288 & ...... & ......  \\
\hline 
\label{tabindex2}
\end{tabular} 
\end{center} 
\end{table*} 

\begin{table*} 
\begin{center} 
\caption{\scriptsize Globular cluster indices errors $\lambda >$4600 \AA.  Missing values are due to limited wavelength coverage. Index errors are derived from photon noise in the unfluxed spectra.}
\renewcommand{\arraystretch}{1.5}
\begin{tabular}{lccccccccccc} 
\hline 
\hline 
ID  &  C4668 & H$\beta$ & Fe5015 & Mg$_{1}$ & Mg$_{2}$ &  Mgb   & Fe5270 & Fe5335 & Fe5406 & Fe5709 & Fe5782 \\
   & (\AA)  & (\AA)    & (\AA)  & (mag)    & (mag)    & (\AA)  & (\AA)  & (\AA)  & (\AA)  & (\AA)  & (\AA) \\
\hline 
  89 &    0.746 &  0.284 &  0.535 &  0.006 &  0.006 &  0.245 &  0.283 &  0.309 &  0.222 & ...... & ......  \\
 124 &    0.539 &  0.186 &  0.393 &  0.004 &  0.005 &  0.179 &  0.199 &  0.224 &  0.165 & ...... & ......  \\
  68 &    0.702 &  0.261 &  0.527 &  0.005 &  0.006 &  0.231 &  0.247 &  0.273 &  0.206 & ...... & ......  \\
 148 &    0.743 &  0.277 &  0.566 &  0.006 &  0.007 &  0.253 &  0.278 &  0.324 &  0.234 & ...... & ......  \\
 175 &    0.595 &  0.224 &  0.453 &  0.005 &  0.005 &  0.207 &  0.221 &  0.250 &  0.183 & ...... & ......  \\
 183 &    0.615 &  0.213 &  0.443 &  0.004 &  0.005 &  0.201 &  0.220 &  0.251 &  0.182 & ...... & ......  \\
 158 &    1.010 &  0.384 &  0.756 &  0.007 &  0.009 &  0.335 &  0.358 &  0.416 &  0.301 &  0.223 &  0.206  \\
 360 &    0.405 &  0.144 &  0.307 &  0.003 &  0.004 &  0.139 &  0.153 &  0.173 &  0.127 &  0.095 &  0.090  \\
 329 &    0.533 &  0.189 &  0.400 &  0.004 &  0.005 &  0.183 &  0.206 &  0.235 &  0.172 & ...... & ......  \\
 277 &    0.734 &  0.272 &  0.558 &  0.005 &  0.006 &  0.239 &  0.266 &  0.302 &  0.223 & ...... & ......  \\
 251 &    0.652 &  0.247 &  0.487 &  0.005 &  0.006 &  0.218 &  0.243 &  0.272 &  0.198 &  0.152 &  0.140  \\
 298 &    0.737 &  0.257 &  0.549 &  0.005 &  0.006 &  0.254 &  0.272 &  0.319 &  0.229 & ...... & ......  \\
 318 &    0.984 &  0.348 &  0.751 &  0.007 &  0.008 &  0.309 &  0.363 &  0.391 &  0.282 & ...... & ......  \\
 606 &    0.406 &  0.149 &  0.311 &  0.003 &  0.004 &  0.140 &  0.152 &  0.166 &  0.123 & ...... & ......  \\
 558 &    0.523 &  0.181 &  0.389 &  0.004 &  0.004 &  0.182 &  0.194 &  0.222 &  0.163 & ...... & ......  \\
 434 &    0.584 &  0.209 &  0.446 &  0.004 &  0.005 &  0.201 &  0.221 &  0.249 &  0.184 & ...... & ......  \\
 462 &    0.591 &  0.222 &  0.455 &  0.005 &  0.005 &  0.207 &  0.227 &  0.265 &  0.190 & ...... & ......  \\
 517 &    0.563 &  0.212 &  0.422 &  0.004 &  0.005 &  0.190 &  0.199 &  0.232 &  0.169 & ...... & ......  \\
 412 &    0.707 &  0.279 &  0.547 &  0.005 &  0.006 &  0.246 &  0.276 &  0.303 &  0.225 &  0.170 &  0.156  \\
 502 &    0.548 &  0.209 &  0.424 &  0.004 &  0.005 &  0.192 &  0.215 &  0.235 &  0.174 &  0.131 & ......  \\
 640 &    0.774 &  0.283 &  0.621 &  0.006 &  0.007 &  0.285 &  0.316 &  0.349 &  0.262 &  0.188 &  0.173  \\
 740 &    0.482 &  0.174 &  0.350 &  0.004 &  0.004 &  0.162 &  0.181 &  0.206 &  0.151 & ...... & ......  \\
 806 &    0.433 &  0.164 &  0.341 &  0.003 &  0.004 &  0.141 &  0.160 &  0.180 &  0.131 & ...... & ......  \\
 899 &    0.428 &  0.154 &  0.322 &  0.003 &  0.004 &  0.144 &  0.163 &  0.183 &  0.135 & ...... & ......  \\
 975 &    0.354 &  0.125 &  0.262 &  0.003 &  0.003 &  0.118 &  0.132 &  0.150 &  0.110 & ...... & ......  \\
1063 &    0.412 &  0.149 &  0.310 &  0.003 &  0.004 &  0.144 &  0.156 &  0.177 &  0.129 & ...... & ......  \\
1011 &    0.566 &  0.213 &  0.444 &  0.004 &  0.005 &  0.203 &  0.214 &  0.247 &  0.179 &  0.135 &  0.127  \\
1037 &    0.530 &  0.206 &  0.414 &  0.004 &  0.005 &  0.185 &  0.198 &  0.223 &  0.169 &  0.125 &  0.115  \\
1145 &    0.446 &  0.164 &  0.330 &  0.003 &  0.004 &  0.150 &  0.168 &  0.190 &  0.137 & ...... & ......  \\
1252 &    0.402 &  0.138 &  0.290 &  0.003 &  0.003 &  0.136 &  0.150 &  0.171 &  0.126 & ...... & ......  \\
1384 &    0.323 &  0.118 &  0.253 &  0.002 &  0.003 &  0.114 &  0.127 &  0.142 &  0.103 & ...... & ......  \\
1126 &    0.607 &  0.203 &  0.455 &  0.004 &  0.005 &  0.208 &  0.230 &  0.262 &  0.193 & ...... & ......  \\
1298 &    0.617 &  0.223 &  0.458 &  0.004 &  0.005 &  0.198 &  0.224 &  0.252 &  0.186 & ...... & ......  \\
1211 &    0.624 &  0.226 &  0.464 &  0.005 &  0.005 &  0.220 &  0.230 &  0.253 &  0.188 & ...... & ......  \\
1098 &    0.686 &  0.255 &  0.510 &  0.005 &  0.006 &  0.226 &  0.258 &  0.299 &  0.221 &  0.158 &  0.155  \\
1182 &    0.608 &  0.232 &  0.456 &  0.005 &  0.005 &  0.197 &  0.226 &  0.245 &  0.186 &  0.133 &  0.129  \\
1574 &    0.483 &  0.174 &  0.369 &  0.004 &  0.004 &  0.166 &  0.185 &  0.213 &  0.150 &  0.114 &  0.106  \\
1443 &    0.692 &  0.263 &  0.504 &  0.005 &  0.006 &  0.230 &  0.280 &  0.311 &  0.218 & ...... & ......  \\
\hline 
\label{tabinderr2}
\end{tabular} 
\end{center} 
\end{table*}

\section{Ages, Metallicities and $\alpha$-Element Abundance Ratios}\label{sec_age}

In this section we describe the measurement of ages, metallicities and
$\alpha$-element abundances which are presented in Table
\ref{tabagez}. We apply the $\chi^{2}$ multi-index fitting technique
of Proctor \& Sansom (2002) for this analysis.  This method involves
the comparison of the measured Lick indices with SSP models (its
application to extra-galactic GCs is described fully in Pierce \etal
2005a,b).  The SSP models of Thomas, Maraston \& Korn (2004; hereafter
TMK04) were chosen because these include the effect of $\alpha$
abundance ratios on the Balmer lines, unlike the models of Bruzual \&
Charlot (2003) and Vazdekis (1999).

We compare the measured Lick indices to the TMK04 SSPs and obtain a
minimum $\chi^2$ fit. This fit is initially sought using all the
indices measured. Simultaneously, a set of $\chi^2$ minimisation fits
are found with each of the indices omitted.  From this set, we select
the fit with the lowest total $\chi^2$, remove the necessary index and
repeat until a stable fit is achieved with no highly aberrant
($>~4~\sigma$) indices remaining. All GCs had some indices that were
significant outliers to the fit and therefore removed during this
process. The errors given for the derived parameters are statistical
1$\sigma$ confidence intervals calculated by a Monte Carlo style
method (see Proctor \etal 2004 and Pierce \etal 2005a for details).

\begin{table*} 
\begin{center} 
\caption{\scriptsize Derived globular cluster properties.  Age, [Fe/H], [E/Fe] and [Z/H] are derived from the $\chi^2$ minimisation process, with errors derived by a Monte Carlo style method.  [Fe/H]$_{BH}$ is derived according to the method of Brodie \& Huchra (1990) from a reduced sample of indices.}
\renewcommand{\arraystretch}{1.5} 
\begin{tabular}{lcccccc} 
\hline 
\hline 
ID     & Age   & [Fe/H]    & [E/Fe]    & [Z/H] & [Fe/H]$_{BH}$ &  Notes\\ 
       &(Gyr)  & (dex)     & (dex)     & (dex) & (dex)       & \\ 
\hline 
  89   &  2.4$\pm$0.6 &  -0.29$\pm$0.15 &  0.36$\pm$0.07 &  0.05$\pm$0.11 & -0.86 & \\ 
 124   &   15$\pm$6.0 &  -1.23$\pm$0.16 &  0.00$\pm$0.20 & -1.23$\pm$0.12 & -1.60 & \\
  68   &   15$\pm$4.4 &  -0.29$\pm$0.13 &  0.12$\pm$0.11 & -0.18$\pm$0.08 & -0.53 & \\
 148   & 14.1$\pm$5.1 &  -0.34$\pm$0.13 &  0.12$\pm$0.10 & -0.23$\pm$0.09 & -1.13 & \\ 
 175   &  2.4$\pm$0.4 &   0.04$\pm$0.11 &  0.12$\pm$0.09 &  0.15$\pm$0.09 & -0.51 & \\ 
 183   &  8.9$\pm$2.7 &  -0.63$\pm$0.12 &  0.32$\pm$0.10 & -0.33$\pm$0.11 & -0.73 & \\ 
 158   &  7.9$\pm$3.6 &  -0.45$\pm$0.21 &  0.18$\pm$0.18 & -0.28$\pm$0.17 & -0.78 & Low S/N\\ 
 360   &  7.1$\pm$2.1 &  -0.87$\pm$0.11 &  0.21$\pm$0.07 & -0.68$\pm$0.11 & -1.32 & \\ 
 329   & 11.9$\pm$2.1 &  -1.47$\pm$0.15 &  0.42$\pm$0.15 & -1.08$\pm$0.09 & -1.55 & \\ 
 277   &   15$\pm$2.7 &  -0.50$\pm$0.12 &  0.24$\pm$0.10 & -0.28$\pm$0.07 & -0.75 & \\ 
 251   &   15$\pm$2.7 &  -0.58$\pm$0.10 &  0.46$\pm$0.08 & -0.15$\pm$0.07 & -0.34 & \\ 
 298   &   15$\pm$6.5 &  -1.14$\pm$0.22 & -0.30$\pm$0.18 & -1.43$\pm$0.15 & -1.74 & NGC 6171 Analog\\
 318   & 12.6$\pm$4.4 &  -1.75$\pm$0.20 &  0.80$\pm$0.14 & -1.00$\pm$0.17 & -1.41 & Low S/N\\ 
 606   &   15$\pm$2.4 &  -0.45$\pm$0.10 &  0.24$\pm$0.07 & -0.23$\pm$0.05 & -0.33 & \\ 
 558   &   15$\pm$6.1 &  -1.02$\pm$0.15 & -0.30$\pm$0.15 & -1.30$\pm$0.11 & -1.25 & Stripped Dwarf?\\
 434   & 12.6$\pm$2.6 &  -0.60$\pm$0.12 &  0.21$\pm$0.08 & -0.40$\pm$0.09 & -0.63 & \\
 462   & 12.6$\pm$3.7 &  -1.03$\pm$0.13 &  0.46$\pm$0.09 & -0.60$\pm$0.12 & -0.90 & \\ 
 517   &  5.6$\pm$1.7 &  -0.46$\pm$0.15 &  0.30$\pm$0.10 & -0.18$\pm$0.10 & -0.70 & \\ 
 412   &  7.9$\pm$3.2 &  -0.57$\pm$0.16 &  0.18$\pm$0.13 & -0.40$\pm$0.16 & -1.07 & \\ 
 502   &  3.0$\pm$0.8 &   0.17$\pm$0.11 &  0.38$\pm$0.05 &  0.53$\pm$0.11 & -0.39 & \\ 
 640   & 13.3$\pm$4.6 &  -2.83$\pm$0.40 &  0.80$\pm$0.38 & -2.08$\pm$0.25 & -2.60 & Very low metallicity\\
 740   &   15$\pm$5.4 &  -1.33$\pm$0.14 &  0.40$\pm$0.10 & -0.95$\pm$0.12 & -1.14 & \\ 
 806   & 11.2$\pm$2.1 &  -0.10$\pm$0.09 & -0.03$\pm$0.11 & -0.13$\pm$0.08 & -0.13 & \\
 899   & 11.9$\pm$2.1 &  -1.25$\pm$0.11 &  0.53$\pm$0.10 & -0.75$\pm$0.08 & -1.13 & \\ 
 975   &   15$\pm$4.7 &  -1.32$\pm$0.09 &  0.42$\pm$0.08 & -0.93$\pm$0.06 & -1.19 & \\ 
1063   &  7.5$\pm$2.4 &  -0.72$\pm$0.13 & -0.09$\pm$0.12 & -0.80$\pm$0.10 & -1.22 & \\
1011   &   15$\pm$4.1 &  -1.01$\pm$0.14 &  0.36$\pm$0.24 & -0.68$\pm$0.18 & -0.66 & \\
1037   &   15$\pm$2.9 &  -0.58$\pm$0.12 &  0.30$\pm$0.09 & -0.30$\pm$0.07 & -0.43 & \\ 
1145   &  8.4$\pm$1.8 &  -0.53$\pm$0.09 &  0.38$\pm$0.06 & -0.18$\pm$0.06 & -0.69 & \\ 
1252   &  8.4$\pm$2.3 &  -1.08$\pm$0.11 &  0.30$\pm$0.10 & -0.80$\pm$0.10 & -1.53 & \\ 
1384   &   15$\pm$5.2 &  -1.26$\pm$0.14 &  0.46$\pm$0.08 & -0.83$\pm$0.12 & -0.95 & \\ 
1126   &  8.9$\pm$2.5 &  -2.18$\pm$0.28 &  0.59$\pm$0.24 & -1.63$\pm$0.19 & -1.63 & \\ 
1298   &   15$\pm$3.5 &  -0.82$\pm$0.12 &  0.42$\pm$0.10 & -0.43$\pm$0.09 & -0.89 & \\ 
1211   &   15$\pm$4.8 &  -1.18$\pm$0.17 &  0.56$\pm$0.19 & -0.65$\pm$0.14 & -0.74 & \\ 
1098   &   15$\pm$2.6 &  -2.00$\pm$0.13 &  0.80$\pm$0.12 & -1.25$\pm$0.08 & -1.72 & \\ 
1182   &  5.3$\pm$1.4 &  -0.07$\pm$0.12 & -0.06$\pm$0.14 & -0.13$\pm$0.09 & -0.73 & \\
1574   & 11.9$\pm$2.6 &  -1.33$\pm$0.14 &  0.40$\pm$0.24 & -0.95$\pm$0.15 & -1.23 & \\ 
1443   &  2.1$\pm$0.6 &   0.57$\pm$0.16 & -0.21$\pm$0.10 &  0.38$\pm$0.17 & -1.06 & Complex metal lines\\
\hline 
\label{tabagez}
\end{tabular} 
\end{center} 
\end{table*} 

The molecular band indices Mg$_1$ and Mg$_2$ are systematically offset
due a number of calibration issues and were excluded for all GCs (see
Proctor \etal 2005; Pierce \etal 2005b). Similar to other GC studies
(e.g. Beasley \etal 2004a) we find the CN indices to be enhanced
relative to the models and therefore they were also removed. 

From our sample of 38 GCs, 8 had poor or uncertain fits.  We will
identify these with open symbols in all figures. In particular, GC 640
is found to be very metal-poor.  We find a Brodie-Huchra (Brodie \&
Huchra 1990) metallicity of --2.6 dex, which is 0.85 dex lower than any
other GC in this sample.  Many of the metal sensitive indices are
below the values spanned by the SSP grids.  One consequence of the low
[Fe/H] is a high [E/Fe] uncertainty.

For GC 1443 there exists a conflict between the indices Fe4531, Fe5270
and Fe5335 which suggest a low metallicity and Fe4383, C4668, Fe5406,
Fe5015 and Mgb which suggest the GC has super-solar metallicity.  We
find the more compelling fit is for a young and super metal-rich GC,
however, this result is uncertain.  Neither the Brodie-Huchra
metallicity nor the photometric colours offer any strong constraint.

Low S/N makes a stable fit difficult for GCs 158 and 318.  The
inclusion or exclusion of single indices, that are well below any
clipping threshold, can change the resulting fit.  GC 318 is
potentially part of the so-called ``H'' GC sub-population noted by
Strader \etal (2005) of intermediate colour GCs, at approximately the
turn-over magnitude.  GC 158 is in the magnitude range as well, but is
too red to be considered a part of this sub-population.

For GC 298, H$\beta$ and H$\gamma$ suggest different ages (similar to
the Galactic GC NGC~6171, see Proctor \etal 2004). The metal sensitive
indices suggest that this GC is metal-poor and therefore more likely
to be older than 5 Gyrs.

The ACS images (Proposal 9401, see Bridges \etal 2006) show GC 558 is
extended and therefore is possibly a stripped dwarf.  The $\chi^2$ fit
for 558 gives an old, metal-poor population with an apparently
negative [E/Fe].  If this really is a stripped dwarf we do not
necessarily expect to be able to fit it with a single stellar population
due to potential stellar population gradients. We note that there are
complex issues with sky subtraction around Mgb for this object on some
nights, therefore Mgb is not included in the fit for this object.

We find two GCs, 517 and 1182, which appear to be of intermediate age
$\sim$5 Gyrs.  Both are around solar metallicity ([Z/H]$\sim$0), but
with differing [E/Fe] (+0.3~vs~--0.06).  It is not clear whether these
GCs are in fact intermediate aged or if the age-metallicity degeneracy
is not broken effectively in these cases.

\begin{figure}
\centerline{\psfig{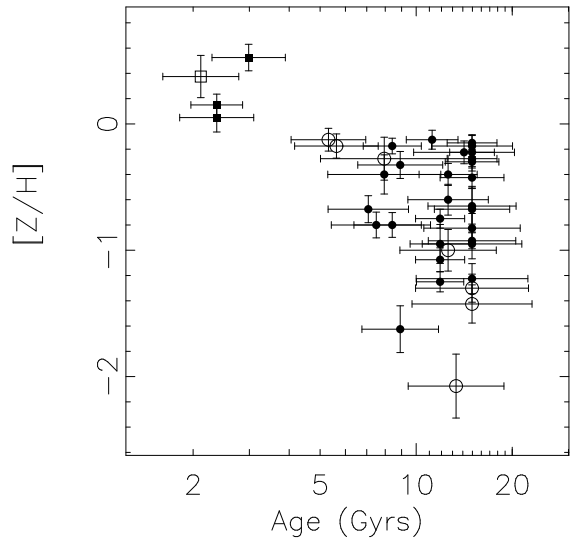}}
\caption{SSP fit metallicities and ages. GCs with uncertain fits are shown as open points.  The majority of GCs are old ($\geq$ 10 Gyrs) with a small number of young, very metal-rich GCs (shown as squares) and several GCs with potentially intermediate ages.}
\label{agez}
\end{figure}

The three clearly young and metal-rich GCs are 89, 175 and 502; these
will be shown as squares on all plots.  Their $\alpha$-element
abundance ratios are all [E/Fe]$\sim$0.3 within errors.  It is
interesting to note that these GCs are not amongst the brightest
objects in our sample.  All three are observed to be photometrically
red.  See Section \ref{sec_disc} for more on the young GCs.

In summary, within errors, we find that the majority of GCs are
consistent with an old age ($\geq$~10~Gyrs).  There are two GCs with
potentially intermediate ages $\sim$5~Gyrs, three GCs that are
definitely young (2-3~Gyrs) and another GC that is most likely young.

\begin{figure}
\centerline{\psfig{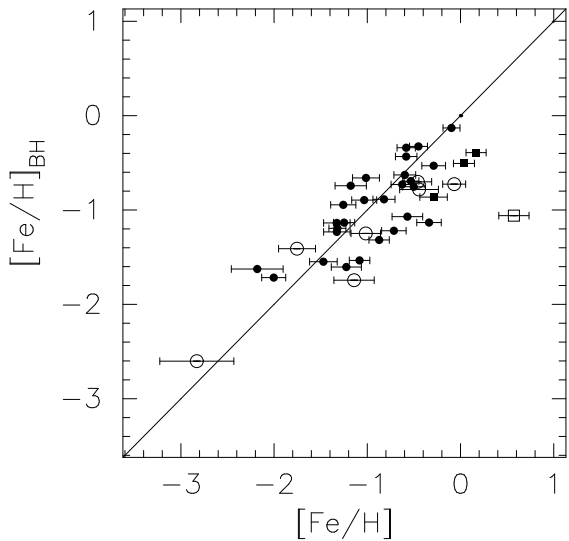}}
\caption{Brodie-Huchra metallicity is shown against SSP metallicity.  The open points are GCs for which the $\chi^2$ fits are uncertain.  This plot shows a general trend of agreement for the majority of GCs. Young GCs are shown as square symbols. The clear outlier is GC 1443 where there are both strong and weak metal indices present in the spectrum.}
\label{febh}
\end{figure}

One means to test our SSP-derived metallicities is to compare them
with those derived by the Brodie \& Huchra (1990; BH) method, which
was originally calibrated to old stellar populations.  Fig.
\ref{febh} shows good agreement for the SSP-derived metallicities of
old GCs and estimates from the BH method.  One would expect the BH
method to give lower metallicities for young stellar
populations. However, the ``young'' objects (ages $<$5~Gyrs), while
systematically offset from the one-to-one line by $\sim$0.5 dex, are
within the scatter of the full sample.

\begin{figure}
\centerline{\psfig{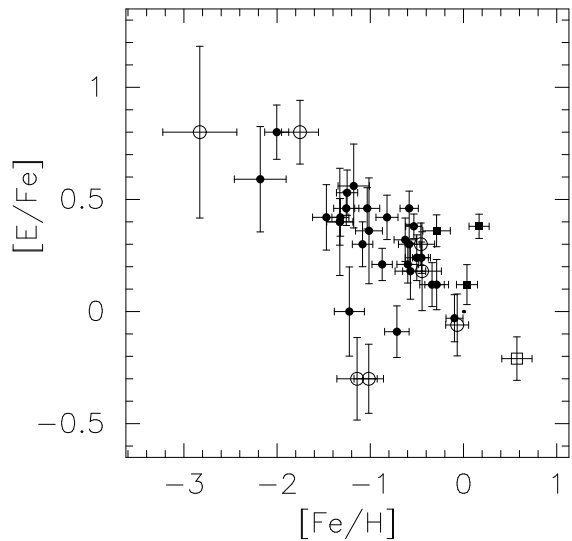}}
\caption{Alpha element abundance ratio vs metallicity.  The open points are GCs for which the $\chi^2$ fits are uncertain and the squares are young GCs.  There appears to be a strong trend of decreasing [E/Fe] with increasing [Fe/H].  However the significance of this trend is highly dependent on the low metallicity GCs ([Fe/H]$<$-1.5) for which [E/Fe] is very difficult to determine.}
\label{fealpha}
\end{figure}

A clear trend of decreasing [E/Fe] with increasing [Fe/H] is seen in
Fig. \ref{fealpha}.  For GCs associated with NGC~3379, Pierce \etal
(2005b) used the index-index plots to confirm the suggestion of a
trend in the $\chi^2$ fit $\alpha$-element abundances. Similar
index-index plots for NGC~4649 GCs do not show the trend that is seen
in the fitted parameters.  In the case presented here there is the
added complication of age differences between GCs, whereas the
NGC~3379 GCs of Pierce \etal(2005b) were found to be universally
old. It should be noted that we find no trend of $\alpha$-element
abundance ratios with age.

\section{Discussion}\label{sec_disc} 

One of the more important factors for the $\chi$ fitting analysis used
in this work is the determination of errors. Uncertainties in the
background subtraction of sky and galaxy light introduces Lick index
errors in addition to the easily quantifiable Poisson noise.  The relative
effect of these errors increases for fainter GCs, so in reality not
only do fainter GCs have larger index errors, but also those errors
are increasingly underestimated.  

Assuming Gaussian errors due to photon noise, 95\% of
measurements should be within 2$\sigma$.  For most objects we measure
$\sim$16 indices, which means that when we apply a 2$\sigma$ clip, only
1 index measurement should be excluded on average per object.  For the
application of $\chi^2$ fitting to extragalactic GCs this is clearly
not the case as we are often forced to exclude 3-7 indices at greater
than 3$\sigma$ before a stable result is obtained.  Here we are also
assuming that the SSP models are perfect.  Both flux calibration and
calibration to the Lick system introduce further errors and
uncertainties.

The main manifestation of these unaccounted for errors is the
increased difficulty in constraining ages and $\alpha$-element
abundance ratios for fainter GCs.  In general, metallicities are less
susceptible to these effects because of the large number ($>$10) of
predominantly metallicity sensitive indices available.  By contrast, there
are only 3 primarily age indicators (i.e. the H$\beta$, H$\gamma$ and
H$\delta$ lines) and only a few strongly sensitive $\alpha$-element
indices.

This underestimation of errors probably occurs in all extra-galactic
GC samples. For example, the large and homogeneous sample of Puzia
\etal (2004) has index errors that appear underestimated by up to a
factor of 2 based on the scatter in index-index plots.  The $\chi^2$
fitting method we apply is immune to this effect as long as the errors
are uniformly underestimated by the same factor across all indices.
However, it is unlikely this is the case as sky subtraction and galaxy
light subtraction could affect every index differently. Despite these
uncertainties, the $\chi^2$ fitting method applied in this work
appears to be robust to problems with individual indices.  GCs
strongly affected by error underestimation are easily identified
during the fitting process.  Modelling errors including horizontal
branch morphology and $\alpha$-element abundance prescriptions are
also factors in the imperfect matching of observed data to the SSPs.

\begin{figure}
\centerline{\psfig{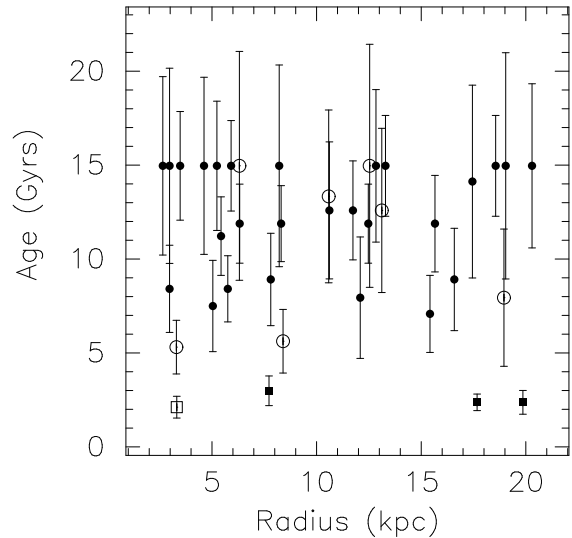}}
\caption{A plot of age vs projected galactocentric radii.   The open points are GCs for which the $\chi^2$ fits are uncertain and the squares are young GCs.  There is no obvious trend with radius.  Of interest are the two young GCs at large projected galactocentric radii $\sim$17--20 kpc.}
\label{radage}
\end{figure}

At least 3, and possibly 4, GCs are found to be young (2-3 Gyrs old)
out of our sample of 38.  This proportion is similar to the 2 young
GCs from a sample of 10 found by Forbes \etal (2001) for NGC~1399,
another large cluster elliptical with no signs of recent star
formation.  A 2-3 Gyr old central burst of star-formation that is
approximately 10\% by mass of the underlying old 10-14 Gyr population
will cause a significant change to the spectral indices of the
galaxy (e.g. Proctor \etal 2005). This has not been reported for the
NGC~4649 galaxy light itself. Indeed, Terlevich \& Forbes (2002) find
the central stellar population of NGC~4649 to be $\sim$11 Gyrs old.
There are no observations to suggest a significant recent burst of
star-formation in NGC~4649 (the UV SED work of Magris \& Bruzual
(1993) suggest minor on-going star-formation).  Do these young GCs
suggest that some of the GCs associated with large cluster gEs are not
formed in that galaxy?  Two of the young GCs (87 and 175) are at large
projected galacto centric radii (see Fig. \ref{radage}).  These GCs
are on the opposite side of NGC~4649 to the nearby companion spiral
NGC~4647 (see Fig. 1 of Bridges \etal 2006).

This result raises several possibilities, firstly that there was
minimal galaxy star formation when these young GCs were formed and
therefore GCs do not trace star formation particularly well. This
would be contrary to what is commonly observed in other
systems. Secondly, that our sampling is not representative of the
overall GC population. This possibility is difficult to rule out.
Thirdly, that GC accretion is a common enough process for large
ellipticals that the GC system is ``contaminated'' by GCs formed in
other galaxies and therefore the GC system is a complex trace of a
combination of star-formation and galaxy assembly. More accurate ages
and $\alpha$-element abundance ratios are necessary to distinguish
between local formation of the young GCs in massive ellipticals or
accretion from nearby galaxies with more recent star-formation.

Our plot of $\alpha$-element ratio with metallicity
(Fig. \ref{fealpha}) is the most visually compelling indication of a
trend of decreasing [E/Fe] with increasing [Fe/H] in the literature to
date.  A similar trend is seen less clearly in Pierce \etal (2005a)
for NGC~1052 GCs, Puzia \etal (2005) for a sub-sample of bright GCs
from 5 galaxies and Pierce \etal (2005b) for NGC~3379 GCs. 

Not all observers find the same trend in $\alpha$-element ratio's.
Strader \etal (2004a) examine NGC~3610 GCs and appear to find an
inverse trend to that presented in our Fig. \ref{fealpha}.  Based on a
comparison of Mgb vs $<$Fe$>$ they find the majority of GCs to have
solar or sub-solar $\alpha$-element ratios.  Four of the metal-rich
GCs appear to have abundance ratios of +0.3 dex; interestingly two of
these are old and two are young.

Olsen \etal (2004), with a sample of 6 GCs associated with Sculptor
group galaxies, find 4 metal-poor ([Fe/H]$<$--1) GCs with abundance
ratios around [$\alpha$/Fe]=--0.3 and 2 metal-poor GCs with
approximately solar abundances (see Fig. 8 of Olsen \etal 2004).  This
is derived from the ratio of Mgb/$<$Fe$>$. We suggest, based on their
Fig. 6, that the calibration to the Lick system of Mgb and Fe5335 for
their data may be a problem.

 It is worth noting that $\alpha$-element abundance ratio's are poorly
constrained for low metallicities and it is therefore possible to
argue that within errors the current larger samples, Pierce \etal
(2005a,b) and Puzia \etal (2005), are consistent with a universal
value of [E/Fe]=+0.3 (consistent with Galactic $\alpha$/Fe measures).
The majority of GCs with well constrained $\alpha$-element abundance
ratio's are relatively consistent with this value within errors.  The
trend seen in all 3 data sets may then be attributed to modelling
effects in the Thomas \etal (2004) models used for all these cases.

A rough order of magnitude estimate can be made of the S/N required to
differentiate the $\alpha$-element abundance ratios of metal-poor GCs.
The expected difference in Mgb between solar abundance and
[E/Fe]=+0.3, for a 10 Gyr [Fe/H]=--1 TMK04 model, is 0.237~\AA.  To be
able to observe a difference of this magnitude spectra with S/N
corresponding to H$\beta$ errors of $\sim$0.13~\AA\ are required.  For
the sample presented here, only two GCs are sufficiently bright to
reach that accuracy.  Calibration to the Lick system and accurate flux
calibration are necessary to make possible the detailed exploration of
$\alpha$-element abundance effects at low metallicities. To completely
resolve the differences in $\alpha$-element abundance ratios currently
measured, spectra of substantially higher S/N than the current samples
are required.

Considering index and index error uncertainties discussed at the
beginning of this section we suggest that further work will be
necessary to confidently determine if the metallicity $\alpha$-element
abundance ratio trend we observe is real or if it is an artefact of
SSP modelling, $\chi^2$ fitting and observation calibrations, both
flux and to the Lick system.

\section{Conclusions}\label{sec_conc} 

We present ages, metallicities and $\alpha$-element abundance ratios
for 38 GCs around NGC~4649 based on our Gemini/GMOS spectra.
These were derived by applying the multi-index $\chi^2$ minimisation
method of Proctor \& Sansom (2002) to the SSP models of Thomas \etal
(2004).  Close agreement is found with metallicity estimates derived
according to the Brodie \& Huchra (1990) method, with the young GCs
systematically offset by a small amount ($\sim$0.5 dex) as expected.
We find 3, possibly 4, GCs with ages of approximately 2 Gyrs, two of
these are at large galactocentric radii (17--20 kpc).  The
$\alpha$-element abundance ratio decreases with increasing
metallicity.

The young GC fraction ($\le$10\%) is consistent with the picture
presented by ``frosting'' models of recent minor star formation in
ellipticals (Trager \etal 2000). In this particular case, noting the
large galactocentric radii of 2 of the young GCs and the lack of
evidence for recent star formation in the galaxy itself, it is quite
possible that the young sub-population was not formed natively, but
instead has been accreted.

\section{Acknowledgements}\label{sec_ack} 
 
We thank the Gemini support staff for help preparing the slit mask.
DF thanks the ARC for its financial support.  SEZ acknowledges support
for this work in part from the NSF grant AST-0406891 and from the
Michigan State University Foundation. This research was supported in
part by a Discovery Grant awarded to DAH by the Natural Sciences and
Engineering Research Council of Canada (NSERC).

These data were based on observations obtained at the
Gemini Observatory, which is operated by the Association of
Universities for Research in Astronomy, Inc., under a cooperative
agreement with the NSF on behalf of the Gemini partnership: the
National Science Foundation (United States), the Particle Physics and
Astronomy Research Council (United Kingdom), the National Research
Council (Canada), CONICYT (Chile), the Australian Research Council
(Australia), CNPq (Brazil), and CONICET-Agencia Nac. de Promocion
Cientifica y Tecnologica (Argentina).  The Gemini program ID is
GN-2002A-Q13. This research has made use of the NASA/IPAC
Extragalactic Database (NED), which is operated by the Jet Propulsion
Laboratory, Caltech, under contract with the National Aeronautics and
Space Administration.

\end{document}